# Memos: Revisiting Hybrid Memory Management in Modern Operating System

Lei Liu, Mengyao Xie and Hao Yang
*State Key Laboratory of Computer Architecture, Institute of Computing Technology (ICT)*
*Chinese Academy of Sciences (CAS)*

**Abstract**
*The emerging hybrid DRAM-NVM architecture is challenging the existing memory management mechanism in operating system. In this paper, we introduce memos, which can schedule memory resources over the entire memory hierarchy including cache, channels, main memory comprising DRAM and NVM simultaneously. Powered by our newly designed kernel-level monitoring module and page migration engine, memos can dynamically optimize the data placement at the memory hierarchy in terms of the on-line memory patterns, current resource utilization and feature of memory medium. Our experimental results show that memos can achieve high memory utilization, contributing to system throughput by 19.1% and QoS by 23.6% on average. Moreover, memos can reduce the NVM side memory latency by 3~83.3%, energy consumption by 25.1~99%, and benefit the NVM lifetime significantly (40X improvement on average).*

## 1. Introduction

In the era of big data and cloud computing, modern computer systems are facing fast growing memory footprint, rapidly increasing energy consumption and high demand for throughput and Quality of Service (QoS). It is critical to have a large capacity memory system that can provide fast data retrieval and operate at a low energy cost simultaneously. Recently, the emerging Non-Volatile Memory (NVM) technologies bring an opportunity to build such hybrid memory systems [12,16,29,43], which can provide DRAM data access speed while having the benefits of low energy, high density, and non-volatility offered by NVM technologies. However, the hybrid memory systems pose new challenges for the memory management mechanism in modern Operating System (OS) [15,33,37,38,39,47,50,54,55,59,60,63].

Conventionally, there are two different ways of organizing a hybrid DRAM-NVM (Fast-Slow) main memory system. In one approach, the faster DRAM is used as a cache of the NVM, which is transparent to OS and user applications [16,44]. Data movement between NVM and DRAM is controlled by dedicated hardware logics. Alternatively, DRAM and NVM can reside horizontally at the same level in the memory hierarchy [16,44], where data placement and movement are managed by software. Compared to the first approach, the horizontal architecture brings more opportunities to optimize the data placement [39,45,53,59,64].

Many studies have discussed memory management mechanisms for "horizontal" hybrid memory system. Compared with them, we have the following insights: (1) To effectively manage a hybrid DRAM-NVM main memory system, it is essential to vertically consider cache activities together with memory channel and bank utilizations through the entire memory hierarchy. The single level memory managing/optimizing approaches (i.e. only cache or DRAM/NVM) may bring sub-optimal performance in many cases. (2) For a hybrid memory system, an effective management scheme should be aware of memory access characteristics at memory page level, since the performance of NVM and DRAM are sensitive to certain memory behaviors such as read/write activities, memory access hotness/hotspot, reuse time, etc. (3) An ideal memory mechanism should try to reduce the memory contentions at the memory hierarchy, and efficiently migrate data among NVM and DRAM for the dynamically changing memory accessing patterns.

In this work, we integrate DRAM and NVM into a Multi-Channel Horizontal Architecture (MCHA), on which memory channels connect to different types of memory. Furthermore, we introduce ***memos***, a memory management framework in OS that can flexibly and efficiently manage hybrid memory systems. We list the key techniques in memos and our design as below:

**(1)** For the first time, we design a full hierarchy memory management framework in OS, vertically scheduling cache, channels, and DRAM/NVM banks simultaneously, balancing the "hotness" to underutilized regions, while reducing the memory interferences at entire memory hierarchy. (Sec. 5)

**(2)** We devise a practical prediction approach that accurately predicts the memory page's future accessing feature based on the history write patterns, guiding the page mapping and migration between DRAM and NVM on the fly in practice. (Sec. 3)

**(3)** We design a kernel-level memory online profiling module (SysMon), which not only efficiently obtains the memory patterns (e.g. memory footprint, page hotness, reuse patterns) at single level (i.e. cache or banks), but also can monitor the memory hierarchy utilization across cache-bank simultaneously. Specifically, SysMon can detect page-level read/write behaviors that are critical for hybrid memory management. (Sec. 4)

**(4)** We devise a cost-effective data migration scheme, which combines DMA and CPU-based page migration approaches together, and adaptively enables the proper approach accordingly. Specifically, we optimize the DMA with a un-lock approach in migration process, thus significantly boosting data transfer rates. (Sec. 6)
**(5)** We implement memos in Linux kernel by modifying the Buddy System, data migration mechanism, and the kernel data structures that denote process and scheduling. Furthermore, we deploy a practical emulation platform for hybrid DRAM-NVM on a real multi-core machine by using the channel-partitioning approach [35]. (Sec. 6)

We test memos by employing diverse workloads, including SPECCPU 2006, Memcached [4] and Redis [9]. The experimental results show that, on average, memos can benefit memory utilization, and contribute to overall system throughput and QoS by 19.1% and 23.6% (up to 28.1% and 34.3%), and outperform the newly proposed approach by around 10%. Moreover, memos can reduce the NVM side memory latency by 3~83.3%, energy consumption by 25.1~99%, and greatly improve the NVM lifetime (up to 500 times).

## 2. Background and New Challenges

Driven by the growing demands for closing the gap between CPU and memory/storage, several NVM technologies emerge as DRAM alternatives, including Phase Change based RAM [24,27,45], Spin-Torque Transfer RAM [29,45], Resistive RAM [18,45,62], etc. These technologies offer the potential of building a low-cost main memory system that has storage level capacity but operates (e.g. read operation) at near DRAM speed. Although these NVM technologies offer unprecedented options and tradeoffs, they do not aim to completely replace DRAM in near future due to its longer write operation latency, higher dynamic energy consumption and even the limited endurance. At present, it is a common wisdom to integrate NVM with DRAM to form hybrid memory systems to mitigate NVM's downsides while leveraging its low leakage and high-density benefits [16,21,40,45,57,64]. The challenges for utilizing hybrid memory are the latency, energy and lifetime differences [24,29,45,61] between DRAM and NVM, showing the **traditional memory management mechanism is not a desirable choice for emerging memory architectures.**

To achieve a desirable and efficient combination of DRAM-NVM, the memory management mechanism needs to be aware of architecture features, memory characteristics and applications' memory access behaviors. Yet, widely used mechanism, such as Buddy System in Linux kernel, manages and allocates physical pages to various applications without these considerations. The physical memory pages and banks selected to serve an application request are nearly random. Prior studies [31,34,36,45] show that this approach can cause significant contention issues at the cache, memory bank and channel level. Some studies attempt to mitigate the problem by using partitioning [31,34,35,37] or vertical memory policies [36,37] that manage cache and bank together. However, **none of these approaches take into accounts all of the memory levels (i.e. cache, channel and bank) at entire memory hierarchy simultaneously (not mentioned in hybrid memory systems)**, leading to sub-optimal memory management and optimization solutions. For example, missing the channel-level consideration can result in channel-level imbalance, interferences/contentions and thus even higher accessing stalls, to which certain type of NVMs can be sensitive. Additionally, the write latency and dynamic energy asymmetry is difficult to be addressed at single memory level as upper levels filter accesses and affect memory read/write patterns at lower levels.

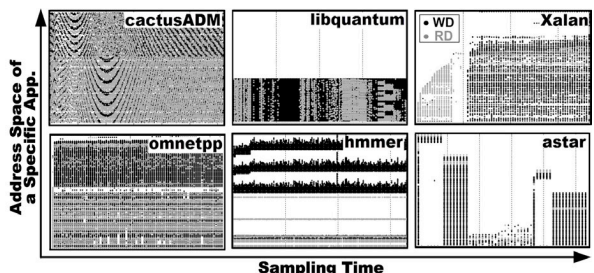

**Fig.1: Write/Read patterns.**

## 3. Discovering Memory Patterns

This section shows our studies on workloads' features, which is critical to hybrid memory management.

### 3.1 Write/Read Patterns

Identifying read/write access patterns for memory pages plays a vital role in selecting DRAM and NVM in respect of the asymmetric read/write access latency. In a specific sampling phase (e.g. Pass in sec. 4.2), we define Read-Domain (RD) and Write-Domain (WD) to indicate whether read or write operations are predominant[1]. Fig.1 shows the RD/WD patterns for several representative applications in SPECCPU 2006 [6]. The horizontal axis denotes sampling time and the vertical axis represents different memory pages in the entire address space. In astar, a large amount of pages exhibit a temporary and transient WD access patterns with the cold access patterns occupying most of the sampling period. In contrast, cactusADM's RD/WD patterns indicate a much more active working set than astar. Most of its pages show a certain level of mix between the WD/RD patterns. These pages should be scheduled to NVM or DRAM dynamically, as placing cold or RD pages onto NVM can save energy cost without significantly harming data access performance, while in other time ranges these pages show relatively intensive write behavior, indicating DRAM will perform better. For applications with distinct memory patterns in different parts of their memory space, it is necessary to segregate the memory space onto different types of memory accordingly. Hmmer and omnetpp are of this type. Generally, Fig.1 tells us, for a specific page, the WD/RD and Hot/Cold patterns may change during run time, and thus a dynamic mechanism is needed to migrate data across NVM and DRAM.

---

[1] We give write operations a heavier weight (empirical value is 2 as write latency is at least 2 times larger than read [16,24,29]). The definitions are: RD: number of reads > 2 * number of writes; WD: 2 * number of writes >= number of reads.

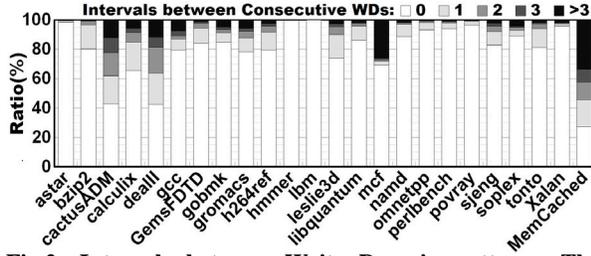

Fig.2: Intervals between Write Domain patterns. The y-axis denotes the rate of intervals between consecutive WDs. Our observations show that for most applications, consecutive WDs are very close to each other.

### 3.2 Predicting Future Memory Features

Furthermore, our comprehensive studies across diverse applications show that, WD usually happens intensively and consecutively. Seen from Fig.2, successive WDs often exhibit very small intervals, and more than 80% intervals between sequential WDs are zero or one. Thus, this inspires us to predict future WD patterns based on the recently observed patterns. However, we are challenged by two questions: 1) how much history information would be used to provide a valid prediction? And, 2) how long can these predictions make sense in the coming future? In order to address these questions, we analyze a large amount of memory trace with WD/RD patterns from diverse benchmarks to reveal the correlation among the number of latest history pattern records, the duration of the future state that can be predicted, and the prediction accuracy. Intuitively, we hope our prediction results can work as long as possible (i.e. the longer the better in x-axis in Fig.3). We try several values of the length of history window (denoted as Window_Len). As shown in Fig.3, in the case where the Window_Len is 8 (i.e. the latest 8 consecutive history records are used in prediction), we can predict a stable pattern in 96% accuracy on average. Fewer records (e.g. Window_Len is 4/6/7) will not provide an ideal prediction accuracy

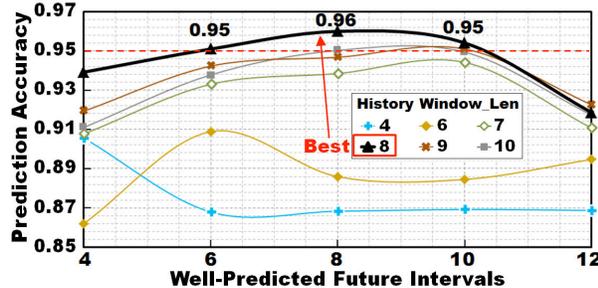

Fig.3: History window and prediction effectiveness. This figure summarizes extensive and diverse cases from all of the SPECCPU 2006 benchmarks and Memcached on average.

Fig.4: Several typical cases of write-history based prediction. We denote WD as 1, 0 indicates RD and cold pages.

(i.e. below 95%). And, more history information (e.g. 9/10) will not contribute to accuracy growth, but only the overhead. Therefore, as a trade-off, we predict the future memory pattern by using the 8 history records. According to the statistics revealed in Fig. 3, a memory pattern predicated by history trace is expected to keep stable for 10 sampling intervals (in x-axis) in 95% cases (accuracy), which is thought sufficiently long to avoid "thrash-out" phenomena caused by mis-prediction.

Fig.4 shows typical examples on how to predict the future memory patterns by leveraging the observed history information. The case_1 denotes a typical WD intensive case, where the page may continuously exhibit high WD frequency in the near future (WD_Freq_H). Case_3 also shows a WD case (Freq_L), but it is not as intensive as case_1. A typical page without WD feature is showed in case_2, and it maybe cold or RD. Finally, in case_4, if the last patterns in K_Len samplings are consecutive WD, we will predict the future as WD state, though it may be marked as Un_WD through the overall view of the sampling window, and visa versa for Un_WD cases. We call this phenomenon *Reverse*. It is a necessary consideration in practice, because sampling windows may span different memory phases. In case_4, *Reverse* means that the sampling window actually span an Un_WD phase and a coming WD phase.

### 3.3 Physical Page-level Reuse Patterns

Beside above, physical page reuse information [37,58] is a relevant element that can determine the scheduling methodology on cache. According to the physical page-level reuse behaviors, we classify physical pages into three categories. The sub-fig.a in Fig.5 illustrates Thrashing pages with very small and stable reuse interval. They may incur severe cache thrashing due to streaming-like memory looking up operations. Allocating large cache resource to them does not improve the system performance [22,36,37]. The sub-fig.b shows Frequently-touched pages with an unstable reuse time and a larger reuse interval, whose data blocks remained in cache may improve cache hit rates [22,36,37]. Finally, the sub-fig.c indicates Rarely-touched pages, which do not necessarily need large cache space. Investigating physical page-level memory patterns is helpful in designing fine grain resource allocation on last level cache (LLC), effecting performance of hybrid memory system.

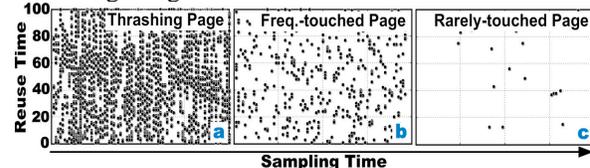

Fig.5: Physical page-level reuse time. The vertical axis of each dot is the reuse time for a sampling at a certain time on horizontal axis. Each sub fig. denotes a sampling for a page.

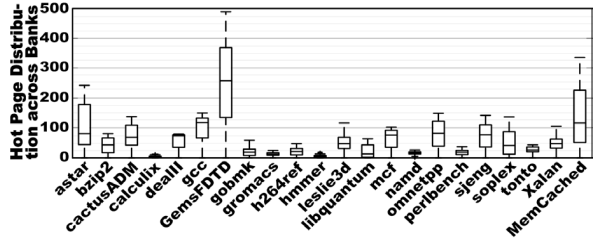

**Fig.6: Bank imbalance[2] on system with 8GB main memory that includes 64 DRAM banks.**

## 4. Monitoring Memory Utilization

### 4.1 Bank Utilization and Cache Activity

In general, modern computer's memory bank system is still the bottleneck of the overall throughput [45]. In order to improve bank utilization for a high memory performance, although common approaches are using physical address interleaving [35,36] and XOR scheme [65] to distribute memory pages as evenly as possible, these approaches don't really play a role in many cases. We find the memory requests distribution over banks are actually quite uneven. Fig.6 shows the standardized distribution of hot (active) pages across memory banks for SPECCPU 2006 and Memcached on average. The larger the variance, the more banks are unbalanced. We can see that applications exhibit more or less bank-level imbalance. GemsFDTD exhibits a very high bank imbalance, which indicates the bank utilization is low. In other words, there are hot banks that suffer severe bank-level conflicts, harming the row-buffer locality and consequently performance. Previous studies only try to distribute the pages evenly, but don't consider the online memory behaviors together, and thus cannot provide ideal bank parallelism, leaving some banks underutilized and lowering the data transfer speed and bandwidth. For hybrid systems, achieving fast data transfer is critical for data migration and rescheduling on both DRAM and NVM. Through rebalancing bank accesses via data migration, underutilized memory banks can share the responsibilities of these "hot" ones, thus contributing to memory utilization, especially avoiding severe bank conflicts in migrating pages between NVM and DRAM.

Moreover, only the bank-level rebalancing is not sufficient. In many cases, although memory accesses are nearly balanced across banks, the cache utilization is still very low. We further study the cache-bank associated address mapping in modern architecture [35,36,45], and find that there are some overlapped bits that index both of the row address in banks and the cache sets. Thus, "blindly" balancing bank utilization without taking into account the data block's corresponding cache address (i.e. row address in a specific memory bank) will lead to cache conflicts. We show a typical example in Fig.7. Suppose there are two groups of pages (part of the row bits are 0010) in different bank0 and bank1. Their data blocks will be mapped into the same cache set group denoted as 0010, thus may cause cache conflicts. If we map a group of pages into other rows denoted as 1011 in Bank1, the conflicts will be eliminated. This motivates us, to pursuit higher memory utilizations, it is essential to consider cache activities vertically together in memory bank allocation and mapping.

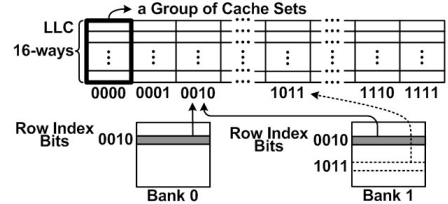

**Fig.7: Cache-bank associated address mapping.**

### 4.2 Inner-OS Memory Page Profiling Module (SysMon)

We open sourced SysMon, an inner-OS profiling module, to collect the above characteristics online (all of the above information is obtained by using SysMon). To determine page hotness, SysMon clears and checks (i.e. sampling) page *access_bit* in PTE in continuous sampling passes. In each pass a given number of samplings (i.e. 100 in default) are performed. If for a page most of samplings detect the *access_bit* being set, the page is actively accessed and thus it is a hot page. Using *access_bit* can also obtain the page-level reuse time and cache utilization [36,37]. To capture WD/RD patterns, SysMon uses *dirty_bit* (also in PTE) and *access_bit* to capture page writes and reads. Similar to the mechanism for detecting page hotness, we examine both *dirty_bit* and *access_bit* in a sampling pass and calculate the weighted ratio of reads and writes. By examining the values of the *bank index bits* in the physical address and counting the pages assigned to each memory bank, SysMon can obtain the bank balance information. Based on the numbers of pages distributed to all the banks, the variation and bank imbalance factor can be calculated. In terms of the cache-bank association, SysMon can obtain the memory hierarchy utilization (details are in Algorithm_ 1). We carefully designed the core data structure for SysMon and use a page shadow array (each element is a raw byte) and bit manipulation to track the memory access patterns. Additionally, by using the PMU [3],

---

[2] The box chart shows the distribution of hot pages for each application. The bottom and top ends on each whisker around a box are minimum (standardized as value 0) and maximum hot page numbers, respectively. The bottom, top, and inner lines of a box are the lower 25% percentile, higher 25% percentile and the median (50% percentile), respectively.

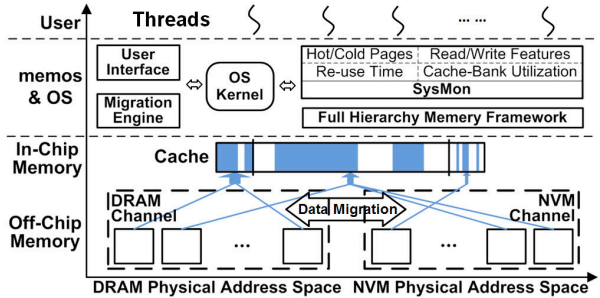
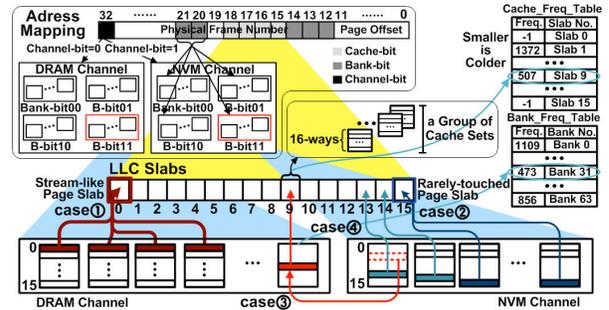

Fig.8: Memos and MCHA in a nutshell.

Fig.9: Address mapping on a typical i7-series CPU [10,33] and memos's overall working process.

SysMon can also obtain per-channel bandwidth utilization information on the fly with a low overhead.

## 5. Memos and MCHA

The design objective of our system is to achieve optimal memory utilization by allocating appropriate type(s) and amount of memory resource according to memory patterns. This section details our design for memos and hybrid memory architecture.

### 5.1 Overview: Managing Hybrid Memory System

Fig.8 shows the overview of our design. On MCHA, the physical address is split into two segments and backed by DRAM and NVM respectively via a specific channel and memory controller (MC). We further design *memos* for the hybrid DRAM-NVM, including memory management framework, data migration engine, and SysMon that captures the runtime information such as hot/cold, RD/WD features and reuse time, etc. Leveraging SysMon, memos first determines data and memory affinity mapping. Then, memos needs to make decision regarding when and how to migrate the pages to another sub-memory system. Specifically, we devise a full hierarchy memory management framework in memos that vertically manages memory resources from the upper levels (i.e. cache), through the middle levels (i.e. channels), to the lower levels of the DRAM/NVM banks at the same time. To achieve a high system performance, memos carefully selects memory medium, banks and even the cache sets for a specific physical page according to its feature on the fly. Finally, for a specific sub-memory system (NVM or DRAM), memos employs bank partitioning and rebalancing approach to reduce bank-level interference.

### 5.2 Full Hierarchy Memory Management Framework

**Fundamental Framework:** We construct the memory management framework by leveraging Page-Coloring approach [31,34,35,36]. As the address mapping illustrated in Fig.9, for a 4K-size page (0~11 bits denote the offset within the page) on a typical 64-bit architecture, bit 32 is used as the channel-bit. Therefore, by selecting a physical page with a specified value (0 or 1) at bit 32, we can control which channel, and consequently which memory segment (DRAM or NVM) to accommodate the page. For cache resource, on our experimental platform, each unique combination of cache index bit values (bits 15,16,17,18 in the page frame denote cache sets index, and also some rows in a memory bank, as shown in Fig.7), or cache-set color, dictates a slab (1/16 of the total LLC capacity) of LLC resource. Thus, we can adjust cache resource allocation by leveraging these bits. Moreover, for the memory bank resource in both NVM and DRAM channel, memos monitors the bank utilization and enables the bank scheduling by using the bank index bits (bits 20,21,14,13 and 12 in Fig.9). Usually, bits 20,21 are used as a combination to uniquely dictate a group of 8 banks (called a bank-group color), and memos can assign additional bank groups by using more than one bank-group colors. Leveraging these bits that index different type of memory resources, memos can form previously unused allocation approaches. For example, there are L Bank-bits, M Cache-bits and N Channel-bits, and memos can use i Bank-bits, j Cache-bits and k Channel-bits to generate an allocation approach represented as (i,j,k) for a group of memory pages, where $0 < i <= L$, $0 < j <= M$ and $0 < k <= N$.

**Channel Allocation:** On MCHA, memory channels are used as a way to designate DRAM and NVM. According to different memory patterns, memos has three allocation principles. Firstly, memos attempts to direct hot pages (e.g. Freq-touched, Thrashing) into DRAM channel, especially for those with WD features. Secondly, RD intensive pages can be directed through

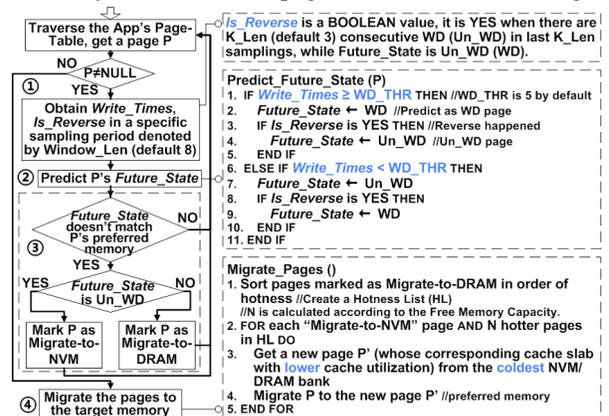

Fig. 10: Migration flowchart and algorithm

```
Algorithm_1: Calculate the frequency of each Bank and Cache Slab
Input: Physical Frame Number (PFN) Output: Bank/Cache_Freq_Table[]
1. clear Bank_Freq_Table[] and Cache_Freq_Table[]
2. FOR each page P IN a specific application DO //calculate frequency
3.   obtain P's bank_id and cache slab_id from P's physical address //PFN
4.   IF P was accessed in a sampling period THEN //by monitoring access_bit
5.     Bank_Freq_Table[bank_id] ← Bank_Freq_Table[bank_id] + 1
6.     Cache_Freq_Table[slab_id] ← Cache_Freq_Table[slab_id] + 1
7.   END IF
8. END FOR
9. RETURN Bank_Freq_Table[], Cache_Freq_Table[]
```

```
Algorithm_2: Get Cold Bank and Cache Slab
Input: Bank/Cache_Freq_Table[]; Output: (1) cold_bank; (2) cold_slab
1. cold_bank ← min(Bank_Freq_Table[])
2. cold_slab ← min(Cache_Freq_Table[]) //except slab 0 and slab 15
3. WHILE rows associated with cold_slab in cold_bank are not free DO
4.   cold_slab ← get the next cold slab in Cache_Freq_Table[]
5. END WHILE
6. RETURN cold_bank, cold_slab
```

NVM channel onto NVM without hurting performance. Thirdly, cold pages are kept in NVM to save energy and reserve DRAM capacity for those hot and WD pages. Moreover, at runtime, memos monitors pages by using SysMon, and migrates them to the proper memory when memory pattern changes.

**Data Migration Mechanism:** Fig.10 highlights memos's migration policies. In step 1, memos records pages' Write_Times in a specific sampling period denoted by Window_Len (default value is 8), and checks whether the WD state is *Reverse* or not according to the last K_Len samplings. In step 2, memos predicts the pages' future state based on the history WD patterns (details in sec. 3.2). Next, memos marks the pages as "will-be-migrated" or not in step 3 regarding the current location and future state. Memos then ranks them in terms of their hotness (i.e. accessing frequency) and puts them into a hotness list (HL). The pages, predicted as WD_Freq_H in future, have a higher migration priority than these Freq_L ones. The real migration will start at step 4 (details are in following section). The entire mechanism works as a loop periodically in reality (default interval is 20s).

In additional, to improve the channel utilization, our design also balances the bandwidth across channels. The key idea is, when the bandwidth bound (i.e. DDR3 is ~7GB per-channel, moving more hot pages to them will not improve bandwidth [35,48]) is achieved as an excessive number of pages being assigned to DRAM, memos will move some RD pages even some WD ones to the underutilized NVM channel. Memos will stop migrating pages from DRAM to NVM when the DRAM channel bandwidth utilization begins to drop. Therefore, the DRAM channel bandwidth utilization is always maximized, and the overall bandwidth is guaranteed to be improved.

**Cache-Bank Associated Allocation:** From the view of memos, LLC is partitioned into 16 slabs (showed in Fig.9) by using LLC index bits (e.g. 15,16,17,18 bits), and each slab denotes a group of LLC sets. Memos employs the Cache/Bank_Frequency_Table in Algorithm_1 to record the utilization of each cache slab and memory bank. The allocation process works as follow: (1) By default, the LLC slabs are further divided into three segments for those three types of pages respectively (in sec. 3.2). All of the Thrashing pages from both DRAM and NVM are tried to map into a small specific reserved cache slab (i.e. slab 0 in Fig.9), isolating them at LLC can help to avoid thrashing other data, especially the data from NVM channel. Meanwhile, all those Rarely-touched pages from both DRAM and NVM channels are mapped together into another reserved slab (slab 15), as usually these pages consume very small cache capacity. Moreover, seen from Fig.9, larger LLC quotas are used for Freq-touched pages from both DRAM and NVM channels. (2) Through iteratively recording each page's accessing times by monitoring *access_bit* in Algortithm_1, memos records the corresponding cache slab and bank utilization in Cache/Bank_Freq_Table. As demonstrated in Fig.9, a lower frequency value means a lower utilization. Algorithm_2 describes how memos obtain the underutilized banks and cache slabs. When moving pages between NVM and DRAM, memos will place them to the underutilized banks (i.e. these lower frequency banks in Bank_Freq_Table) for better bank parallelism, thus avoiding bank conflicts caused by blind mapping. Simultaneously, by placing pages to the rows whose index bits are associated with the low utilization cache slabs in Cache_Freq_Table, data can be loaded to these underutilized cache slabs. Doing so, memos can help to improve both cache and memory bank utilization while reducing the memory conflicts in those "hot" regions at memory hierarchy. (3) If the associated memory regions (i.e. rows) in target bank in (2) are not free, memos will try to select another underutilized slabs in Cache_Freq_Table accordingly, whose associated rows are still in this bank. If the memory banks in DRAM channel cannot provide sufficient capacity, memos will just migrate $N = \sum_{i=0}^{BANK-1} \sum_{j=0}^{ROW\_GROUP-1} (FMC_{ij} / Page\_Size)$ pages with higher migration priority in HL, where $FMC_{ij}$ denotes **f**ree **m**emory **c**apacity (FMC) of rows in $j^{th}$ row_group (corresponds to $j^{th}$ cache slab) within $i^{th}$ DRAM bank. (4) Memos will enlarge the reserved slabs, if the associated memory capacity cannot meet the special requirements (e.g. stream-like application with large memory footprint).

**5.4 Cases on Overall System Working Process**

To better understand the overall scheduling process, Fig.9 shows several typical cases. As illustrated in case 1, pages that exhibit stream-like patterns are mapped into cache slab 0, meanwhile they are distributed into different memory DRAM banks for better bank-level parallelism (i.e. 15~18 bits, denoting both the row and cache set index, are with the value of 0). Similar things happen in cache slab 15, which is reserved for Rarely-touched pages, especially for these pages with relative lower access frequency and kept in NVM side (case 2). For NVM, ideal bank-level parallelism can hide the expensive access latency (sometimes raised by memory interferences), as NVM systems often provide a large amount of banks. In case 3, memos migrates a hot/WD page from NVM to DRAM across channels. It firstly selects a coldest memory bank (for higher bank parallelism and overall utilization), and then maps the page to the corresponding row that associated with proper cache slab with a lower utilization (slab 9 in Cache_Freq_Table). Moreover, NVM channel can also provide bandwidth. As shown in case 4, data blocks from RD pages can be loaded to cache directly through NVM channel. Note that for these RD pages with Thrashing feature in NVM, memos will also map them to the reserved slab 0. Even for a specific channel, hot pages are migrated from highly utilized banks to lower ones to balance the overall utilization.

**Table 1: The parameters of NVM, DRAM and Cache [16,37].**

| L1 cache | 32KB instruction cache, 32KB data cache, 64B cache block |
|---|---|
| L2 cache | 256KB data cache, 64B cache block |
| L3 cache | 8MB data cache, 64B cache block |
| DRAM system | $t_{RCD}$=10ns, $t_{RP}$=10ns, $t_{WR}$=10ns, endurance=N/A, read (write) energy=51.2 (51.2)nJ, standby power=1W/GB |
| NVM system | $t_{RCD}$=20ns, $t_{RP}$=23ns, $t_{WR}$=160ns, endurance=$10^6$, read (write) energy=102.4 (512.0)nJ, standby power=0.1W/GB |

## 6. Emulation and Kernel Modules

In this section, we show the implementation details. We implement memos in Linux kernel with the version 2.6.32.15. As NVM is not available in market, we set up an emulation environment for the NVM evaluations.

### 6.1 Deployment of MCHA

We emulate MCHA (in Fig.11) by deploying a channel-partitioning mechanism [35] to divide the memory address space into two segments on a server with Intel i7-series CPU and dual-channel with DDR3 memory[3]. We mark the channels as DRAM and NVM channel respectively, and set up a PIN-based simulation for NVM channel by employing modified DRAMSim2 (with NVM parameters) [1] to simulate the NVM performance (i.e. energy, and latency). As PIN tool [5] gets the memory traces with cache behaviors, we further use a cache simulator, DineroIV [2], to filter the memory accesses on cache hierarchy. Table 1 shows the parameters in detail. Additionally, we employ PMU [3] on each memory channel, and thus we can get more details on cache miss and bandwidth utilization. Memos runs smoothly on this emulated MCHA. Fig.11 further gives an example of scheduling 1/16 cache to NVM channel by mapping data vertically to physical pages, whose 15~18 bits are 1 in blue color.

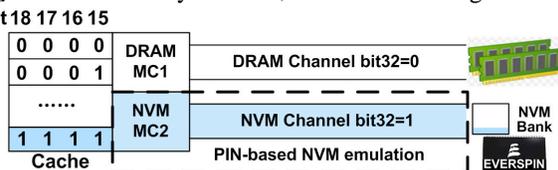

**Fig.11: Illustration of the MCHA emulation platform.**

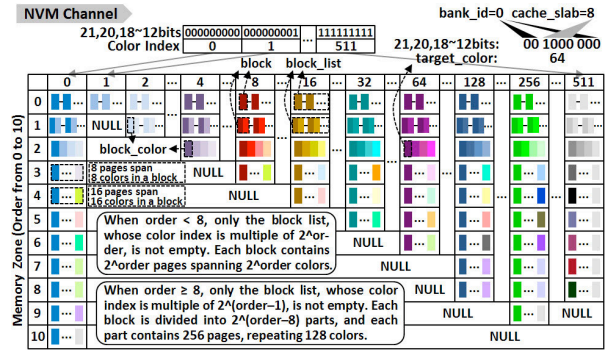

**Fig.12: Sub-buddy System for NVM.**

### 6.2 Supporting Memos in OS kernel

To implement the core of memos' memory framework, we extend Page-Coloring in Linux kernel to reorganize free pages in Buddy System by using the channel, bank and cache bits in Physical Frame No. (PFN) simultaneously. With channel bit, we reorganize all physical pages into two sub-buddies logically, one for pages in NVM and the other for pages in DRAM. In each sub-buddy, we can still use other index bits (cache/bank bits) to allocate resources. By doing so, memos actually tags resources according to hierarchy details, material features, and therefore can efficiently allocate them accordingly. Fig.12 shows the organization of sub-buddy for NVM channel. Nine bits (21,20,18~12 bits) in PFN form a set of 512 colors (000000000~111111111). In order 0, each block only has one page and the block list under a particular color is a set of pages of that color. For example, the block list indexed by color 1 (i.e. light blue block in Fig. 12, index bits 000000001) in order 0 contains any free and non-continuous pages with color 1. Order 1 is similar to order 0 except that each block has two continuous pages, and thus block list under color 1, 3, 5 and etc. are empty. In order 1 list, two pages in each block span two colors. Other orders are organized in a similar way.

**Algorithm_3:** Page Allocation Hashing Algorithm
**Input:** (1) order; (2) target_color **Output:** one page with target color
//CASE: Physical pages organized based on bits 12~18, 20~21
1. **SWITCH** (order)
2.

| order | 0 | 1 | 2 | 3 | 4 | 5 | 6 | 7 | 8 | 9 | 10 |
|---|---|---|---|---|---|---|---|---|---|---|---|
| colors_per_block | 1 | 2 | 4 | 8 | 16 | 32 | 64 | 128 | 128 | 256 | 512 |

3. **END SWITCH**
4. block_color = (target_color / colors_per_block) × colors_per_block;
5. page_index = (target_color − block_color) % 128 +
   ((target_color − block_color) / 128) × 128
6. Expand_color_block (page_index, order)
7. **RETURN** page[page_index] and remove this page from Buddy

* target_color is the color of the requested page.
* block_color is the color of the first page in a block.
* colors_per_block is the number of colors in a block.

---

[3] We try different memory configurations on Multi-Channel Horizontal Architecture (MCHA). The memory capacity in NVM channel ranges from 4 to 16GB (usually larger), and the DRAM is 4GB. Fig.14 shows the experiments.

Memos uses algorithm_3 to allocate a page with O(1) time consumption and will be O(log(n)) when memory blocks are frequently merging and splitting. An example is showed in Fig.12, a specific page corresponds to cache slab 8 and bank 0 in NVM channel) can be obtained in order 2 list with color 64.

The primary memory allocation interface in kernel is *alloc_resource (int channel_id, int cache_slab, int bank_id)*, which is used to obtain a group of memory resources. By adding resource control parameters into task_struct (denoting *Process* in Linux), users can leverage this interface to map the application's data stack according to their requirement.

### 6.3 Data Migration Engine

To provide a high efficient data migration, we design and implement a new data migration engine, which combines CPU and DMA-based page migration simultaneously. For the CPU approach, based on the page copy primitive provided by OS kernel, we implement a lock involved page migration mechanism. This approach needs locking the page before migration, ensuring the data are consistent after the migration. Besides, we devise a DMA-based unlocking migration approach, which is quite efficient for large memory blocks. In this approach, data are migrated through DMA channel first (note that we do not lock these pages before/after migration), and then we check whether these pages are modified during the migration by checking the *dirty_bit* in PTE when migration finishes. We then create the new PTEs for these successfully migrated pages, whose *dirty_bit* is 0 (i.e. not modified during migration), and discard those dirty pages. The migration engine works iteratively. In practice, the DMA approach works better in the DRAM to NVM migration cases, where most of will-be-migrated pages are usually cold or RD ones, which are not modified during migration period.

We further develop a suite of interfaces that expose the page migration mechanism to memos, SysMon and the outside users. By leveraging them, we implement two migration approaches, lazy and eager, to support data transfer across and inner memory channels. By default memos enables lazy page migration, which moves pages when necessary, and leaves eager approach (immediately move page upon requests) to users. We list some of these interfaces. The interface *migrate_cpu (struct page * src, struct page * des)* is evoked when memos moves a specific page to its destination, and can be used iteratively. Usually, memos enables it in the cases where moving a small number of hot and WD pages to DRAM. On the contrast, with Scatter-Gather [8] mode, after the DMA initiation, DMA based migration approach iteratively uses the interface *dma_memcpy_pg_to_pg (dma_channel, oldpage, newpage)* to move pages, thus

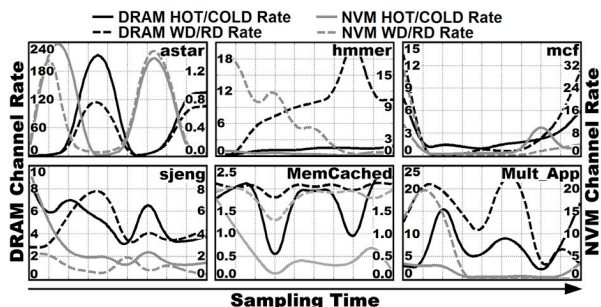

Fig.13: Effectiveness of memos in resource scheduling.

it can effectively move a large number of physical pages with discrete addresses.

## 7. Evaluation

We evaluate memos by using a wide range of workloads including applications from SPECCPU 2006 [6], and the widely used Memcached [4] with data from Twitter in 8GB footprint and Redis [9].

### 7.1 Effectiveness of Managing Hybrid Memory

Memos can capture memory access patterns and migrate data pages accordingly on the fly. Fig.13 shows the effectiveness of memos based on HOT/COLD rate that denotes the ratio of hot pages to cold pages, and WD/RD rate that shows the ratio of write operations to read operations (at page level) on DRAM and NVM, respectively. In our experiments, we map applications to NVM at the beginning, as NVM might be used as hard disks in practice, and most required data are moved from NVM into fast memory on demand.

Taking hmmer as an example, DRAM HOT/COLD rate increases stably over time, indicating that hot pages are migrated to DRAM continuously. Meanwhile, DRAM WD/RD rate shows a similar but a sharper increasing trend. By contrast, HOT/COLD and WD/RD rates in NVM exhibit consistently declining trend, illustrating that pages with RD features and relative low memory access frequency are moved to NVM. The rate curve often fluctuates, shown in astar, as it may exhibit diverse and dynamically changing memory access behaviors. The bottom-right subfigure shows the metrics for a multi-programmed workload that includes several SPEC applications. We observe that memos can handle the diverse memory access patterns well by segregating pages into different memory sub-systems based on memory access frequency and write/read patterns. We further test Memcached, whose active working footprint is small but changes frequently. Fig.13 shows memos can handle these changes by migrating hotspots into DRAM (fast memory) to benefit the overall performance, and therefore the DRAM side HOT/COLD rates are higher than these in NVM channel. Memos also tries to distinguish and segregate WD and RD pages across different channels, and thus the WD/RD rate is always higher in DRAM channel, though Memcached often exhibits very unstable write/read features. Generally, the overall HOT/COLD (85.4%) and WD/RD (83.2%) rate in DRAM channel is larger than these in NVM due to the effectiveness of memos on moving hot/WD pages to DRAM, while keeping cold/RD pages in NVM. Doing so will also brings following benefits.

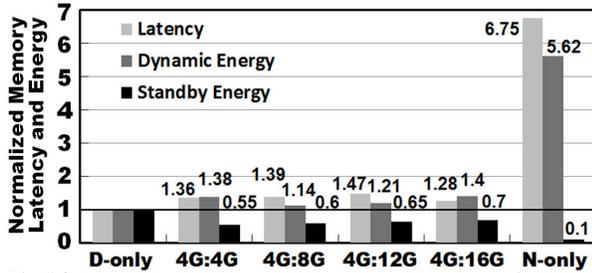

Fig.14: Memory latency and energy comparison on different memory architectures. D/N-only is short for DRAM/NVM-only system. The x:y denotes DRAM and NVM capacity of different configurations on MCHA.

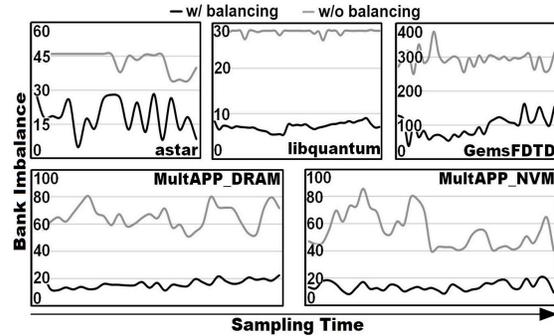

Fig.15: Bank parallelism improvement (lower is better).

**Energy and Memory Latency Reductions:** These experiments[4] are carried out on the emulation platform in Fig.11. We calculate the average value of the typical phases of memory traces. For mcf, the dynamic energy in NVM channel is significantly reduced from 2.1 mWatt to 0.001 mW, while the average memory access latency is reduced from 102.0 ns to 55.7 ns. For xalan, the energy reduction is from 1.2 mW to 0.3 mW, and the memory latency reduces from 78.0 ns to 61.2 ns. For computation intensive applications, which do not issue as many memory requests as memory intensive applications, our approach also brings benefits. For instance, hmmer's energy consumption is reduced from 0.7 mW to 0.5 mW, and its memory latency is optimized from 68.0 ns to 62.6 ns on average. For the widely used Memcached, the energy consumption is reduced from 0.2 mW to 0.1 mW, while the memory latency is also improved to 60.0 ns from 62.0 ns, on average. For Redis, the memory latency in NVM channel is reduced from 62.9 ns to 58.5 ns, and the dynamic energy is reduced from 0.7 mW to 0.2 mW.

The overall performance is showed in Fig.14. Generally, in terms of memory latency and dynamic energy consumption, memos on MCHA (DRAM-NVM) exhibits a much better performance than NVM-only cases (79.6% and 77.2% reduction for memory latency and dynamic energy, respectively), though it may bring a slight higher dynamic energy consumption and latency than DRAM-only cases. But if we consider the overall standby energy, memos on MCHA may be a more cost-effective approach than the DRAM-only. Fig.14 also shows memos' scalability. Memos performs well in the cases where memory capacity in NVM channel ranges from 4 to 16GB (the DRAM is 4GB).

**NVM Lifetime Improvements:** For lifetime calculation, we model the NVM memory with cell endurance of Endurance_X ($10^6$ in Table 1). The NVM memory is operated at memory block granularity (i.e., 64 bytes). Also, we assume the NVM memory uses an effective write leveling scheme (e.g., Start-Gap [25]), thus the overall NVM manages to achieve an overall lifetime which is 95% of the average NVM cell lifetime. Experimental results show that memos on MCHA can random memory channel interleaving scheme, on improve the NVM life by 40X (up to 500X) against average. For example, in single application case, where only hmmer is launched and runs in a long time, the NVM lifetime is only 3.2 years. Yet, on the MCHA platform with memos, the corresponding lifetime is improved to 108.8 years. Another case is mcf, in which the NVM lifetime is merely 0.17 year, but achieves 73.3 years by using our optimizations. We further test a multi-programmed workload (including hmmer, Xalan and mcf), which exhibits memory intensive behaviors and issues a large amount of write operations. Our results show that the NVM lifetime is only 0.14 years in common cases, while on the contrary, the lifetime achieves 44.7 years on MCHA with memos employed.

These improved results are from: 1) the number of write operations is significantly reduced in NVM channel, as WD hot pages are largely moved to DRAM; 2) memos tries to reduce the memory contentions on cache and banks, and thus improves the memory hierarchy utilization. All of these contribute to NVM lifetime, energy saving and latency reduction.

**7.2 Rebalancing and Caching Effects**

As mentioned before, bank imbalance hurts the overall system performance. To address this issue, memos rebalances hot pages across memory banks for NVM and DRAM channel by mapping the migrated pages to the current coldest bank across channels and partition banks among threads [34], contributing to the reduction of bank conflicts and bank balancing. In single thread cases, seen from Fig.15, by rebalancing hotness across banks, the imbalance (measured by standard deviation of the number of active pages between hottest and coldest banks) is significantly reduced by around 60~70%. In multi-programmed cases, seen from the Multapp_DRAM/NVM sub-figures, the standard deviation of bank imbalance drops to around 20 stably in both DRAM and NVM channel. These experimental

---

[4] Baseline is interleaved mapping scheme across channels, and page interleaved mapping across banks inner a channel [34,65].

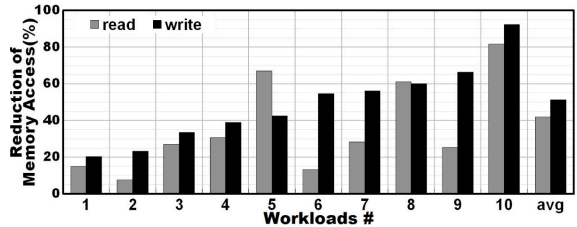

Fig.16: NVM channel memory access reduction.

results indicate that memos works well in memory bank balancing, benefitting the bank utilization.

Besides, by carefully placing data in a specific bank in migration process, memos can map the data blocks into the cache slabs in a lower utilization, while the reserved slab for the Thrashing pages can help to eliminate the cache contentions. On average, the overall LLC misses are reduced by 27.4% compared to commonly used cache-hashing mapping [10,37]. Notable, in addition to moving WD pages to DRAM channel, the write operations in NVM side are further reduced by around 50%, while read operations are reduced by 42% across 10 randomly generated SPEC workloads. The maximum benefit is in workload 10, where the thrashing pages are mapped to DRAM channel and the corresponding cache quota is minimum (1/16 of the entire cache, 512KB on our platform), leaving the rest space to NVM channel. LLC actually works as a filter to cache a lot of expensive dynamic operations to NVM, especially the write operations.

Mentioned before, as memos segregates pages into different channels, in some cases, it might lead to imbalance of bandwidth utilization among channels. If users want to improve the overall bandwidth utilization, memos can migrate hot pages (even WD ones) to NVM channel for bandwidth rebalancing, as long as DRAM channel bandwidth is not harmed. Our experimental results over 102 workloads show, by comparing the unbalancing cases, memos can reduce bandwidth discrepancy by 24.6% between DRAM and NVM channel on average, indicating that the overall bandwidth usage is improved by around 20%.

### 7.3 Overall Performance of Memos on MCHA

We compare memos on MCHA with some typical resource scheduling approaches, i.e. cache-bank vertical management (w/o channel optimization used compared with our new approach) [36,37] and utility-based cache partitioning [31]. Due to the limited availability of NVM DIMMs, we use DRAM in all channels. To mimic the real world workload patterns, we launch several randomly selected applications to run together, and keep injecting applications at different times. Those previously launched applications may terminate upon completion. Fig.17 illustrates the results. Memos with MCHA achieves an average of 19.1% performance gain, and outperforms the previous best approach by 7.3% (up to 11%). Memos performs well on most of the sampling points, especially points 6, 8, 10, 11, and 16. We further find that at these time points pages with streaming accesses from thrashing applications (e.g. libquantum) are well confined into a dictated memory channel with a specific LLC slab, **thus minimizing the interferences among the applications with conflicting behaviors in both channel and cache**, while the bank utilizations are nearly balanced. At point 16, memos achieves the highest performance gain (around 28%), due to Thrashing and Rarely-touched pages severed by different channels with a small LLC quota respectively, while frequently accessed pages from all channels are allocated to use a larger amount of LLC, and the channel-level bandwidth utilization is nearly balanced. Thus the memory interferences are greatly reduced across entire memory hierarchy (including channels). Moreover, memos with MCHA can improve the overall QoS (indicated by Max Slowdown) [34,36,45] significantly (23.6% on average). Take the case in point 11 as an example, memos on MCHA can bring 34.1% benefits for QoS, while the utility-based cache, and vertical cache-bank partitioning with dual-channel interleaving scheme improve QoS by 13.2% and 19.4%. Our new mechanism exhibits obvious advantages, and outperforms them by 20.9% and 14.7%, respectively. Memos doesn't perform as well on certain points (e.g. points 1, 2 and 4), where memory footprint and bandwidth requirements are relative low. This is mainly because of the overhead caused by balancing, sampling and migration offset the benefits.

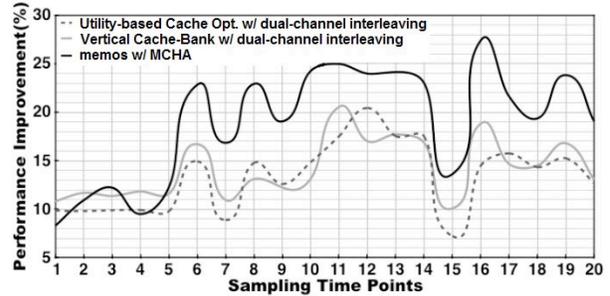

Fig.17: Overall performance throughput improvement[5].

However, in general, memos performs well in memory intensive and high interference cases. Because the **full hierarchy memory optimization mechanism reduces the memory interferences over the memory hierarchy (i.e. cache, channel and memory banks)**, outperforming the vertical cache-bank (without channel-level consideration) and single-level cache optimizations. We believe that our results can reflect the performance for hybrid DRAM-NVM systems, as we show that a predominant portion of the write operations 83.2% on average are moved to DRAM, while NVM is primarily used for reads, which have similar latency on DRAM and NVM. Moreover, our experiments show that with carefully data mapping across the memory hierarchy, the computers equipped with hybrid memory can achieve similar or even better overall performance than the ones with all DRAM.

---

[5] The baseline is the unmodified Linux kernel. The three schemes perform differently against the general baseline.

### 7.4 Overhead

Memos's overhead comes from the following sources. 1) The cost of page table traversal used in SysMon. It depends on the workloads' memory footprint in practice, and can be reduced by increasing sampling interval with an increasing step once it collects sufficient information. Doing so will significantly reduce the overhead for long running workloads with relative stable memory patterns. Moreover, for workloads with an extremely large memory footprint, random sampling can be adopted. Our studies show sampling 5~15% pages periodically can cover the entire memory space after several sampling passes. 2) Page migrations. Migrating a 4KB page costs 3us on our platform in CPU migration cases. The average overhead is less than 8% on average in our experiments with lazy migration. Moreover, with our new DMA migration engine, the overhead can be further reduced.

## 8. Related Work

**1) Optimizing Throughput/QoS using OS Approaches.** To provide ideal performance and QoS, the work in [13,51] design data/control plane for OS, and isolate the crucial path from kernel operations. Some studies [19,28,32,46,52] focus on application and architecture-aware memory polices and task management on multi-core servers. Our work distinguishes with them on full hierarchy memory policies by considering the memory hierarchy across cache, memory channel and banks together. Thus, memos can bring additional performance benefits over previous work, and an even larger optimization space. **2) Designing/Optimizing Hybrid Memory System.** Many efforts design the new hybrid memory architecture [12,16,23,29,43], and typical work in [27,61,67] optimize the memory controller logic, write operations, and row-buffer locality for NVM performance. Some work further study the memory management accordingly [15,17,20,26,39,50,53,54], and even for big data and virtualization environments [49,56,63]. The latest studies redesigned systems (e.g. databases) for NVM used platforms [11,41,66] and end client devices [30]. **3) NVM Allocator.** The work in [7] provides an open source NVM allocator for both persistent and volatile usage. SSDAlloc [14] provides API to users for using SSDs on hybrid memory systems. Our work is complementary to these efforts in 2) and 3). We provide a memory framework in OS to handle diverse memory requirements on hybrid fast-slow memory system, and can be easily deployed. **4) Page-Coloring based Memory Partitioning.** Many previous studies [31,34,36] use cache/bank indexing bits in physical address mapping scheme to partition memory resource for performance. Our work is different from them on the involved address bits, including not only the cache and bank bits, but also the channel bits simultaneously. **5) Page Access Monitoring.** Many previous work [22,38,50,53] conduct online profiling by leveraging hardware performance counters. In this paper, SysMon can obtain the memory hierarchy utilization on the fly at OS level without hardware supports. Specifically, SysMon can detect page-level reads/writes, which are critical to consider in hybrid memory environment.

## 9. Discussion

**1) Portability.** Memos relies on memory indexing bits, which can be obtained from venders' manual or by employing detection approaches in [34,35]. In practice, we parameterize the architecture features (e.g. number of channels/banks) and other thresholds/parameters (i.e. WD_THR, K_Len, sampling/migration interval .etc) as inputs to memos. Therefore, memos can be easily ported to platforms with diverse configurations. **2) Usages in Reality.** Memos provides interfaces to users. If environment changes, users can profile training sets and determine the running parameters for production workloads. Pre-profiling for system optimization is commonly adopted [42], and has been practiced in Google datacenters. What's more, in our design, users are expected to enable thread/task memory mapping to pin some applications with very stable or frequent changing memory patterns in their preferred memory, if these workloads' memory features could be obtained by pre-profiling approaches (SysMon and hardware performance counter can be used in such methods). **3) Memory Footprint.** We tested memory footprint from a few MB to 8 GB, which we believe cover most memory footprint scenarios for personal and enterprise applications. For example, VMware's vCenter Server, which is considered relatively heavyweight, consumes around 8 GB memory on a single machine. Enterprise software venders are reducing the memory footprint of a single software component and scaling the system across multiple machines. Moreover, memos would be practical in big data cases where Spark and Hadoop are involved by mapping these RD stream-like pages (from NVM side) into a small quarto in the cache to facilitate the overall performance.

## 10. Conclusion

This paper designs memos to meet the challenges on the platform with hybrid memory system. To achieve a high overall performance and ideal quality of service, memos schedules the resources at the entire memory hierarchy according to memory patterns and NVM/DRAM's features. Memos can be deployed on systems equipped with the Fast-Slow memory system.